\documentclass[twocolumn,english,aps,prb,floatfix,showpacs]{revtex4}
\usepackage{times}
\usepackage[T1]{fontenc}
\usepackage[latin1]{inputenc}
\usepackage{color}
\usepackage{amssymb}

\makeatletter

\usepackage{graphicx}
\usepackage{bm}

\usepackage{babel}
\makeatother
\begin{document}

\title{Entanglement between static and flying qubits in quantum wires}

\author{J.H. Jefferson$^{\textrm{1}}$, A. Ram\v{s}ak$^{\textrm{2}}$, and
T. Rejec$^{\textrm{2,3}}$}

\affiliation{$^{\textrm{1}}$QinetiQ, Sensors and Electronics Division, St. Andrews
Road, Great Malvern, England}

\affiliation{$^{\textrm{2}}$Faculty of Mathematics and Physics, University of
Ljubljana, J. Stefan Institute, Ljubljana, Slovenia}

\affiliation{$^{\textrm{3}}$Ben-Gurion University, Beer Sheva, Israel }

\date{12 August 2005}

\begin{abstract}
A weakly bound electron in a semiconductor quantum wire is shown to
become entangled with an itinerant electron via the coulomb interaction.
The degree of entanglement and its variation with energy of the injected
electron, may be tuned by choice of spin and initial momentum. Full
entanglement is achieved close to energies where there are spin-dependent
resonances. Possible realisations of related device structures are
discussed. 
\end{abstract}


\maketitle
A major goal in the rapidly emerging field of quantum information
processing is the controlled exchange of quantum information between
propagating and static qubits. Purely electron systems have potential
as entanglers due to strong Coulomb interactions and although charge-qubit
systems suffer from short coherence times, spins in semiconductor
quantum wires and dots are sufficiently long-lived for spin-qubits
to be promising candidate for realizing quantum gates involving both
static and propagating spins \cite{Loss98,Elzerman04,Hanson03,Hanson05,Petta04}.
Entanglement between propagating electron pairs has been proposed
using an electron beamsplitter\cite{burkard04}, a double-dot electron
entangler exploiting the singlet ground state \cite{hu04}, the exchange
interaction between conduction electrons in a single dot \cite{costa01,oliver02}
and the exchange interaction between electron spins in parallel surface
acoustic wave channels\cite{Barnes00}. In this letter we propose
a scheme whereby a single propagating electron interacts strongly
with a bound electron in a quantum wire. This differs from the quantum-dot
systems referred to above in several respects. Firstly, entanglement
is induced between the spins of one propagating and one bound electron,
rather than two propagating electrons, and this entanglement is detected
directly by measuring electron spin, rather than indirectly through
current-current correlations. Secondly, the entangling interaction
between the propagating and bound electron in the quantum wire is
enhanced compared with a quantum-dot system, giving rise to spin-dependent
resonant bound states that are a consequence of the Coulomb interaction
and electron antisymmetry, rather than externally imposed barriers.
This allows considerable flexibility in controlling the entangling
interactions via the kinetic energy of the incident electron.

Consider a semiconductor quantum wire in which there is a weak confining
potential which is capable of binding one, and only one, electron.
Slight deviation from a perfect 1D confining potential, either accidental
or deliberate, can give rise to fully bound states for electrons.
When the confining potential is very weak, such as occurs with a weak
symmetric bulge in an otherwise perfect wire, there is one and only
one bound state \cite{1bound}. Furthermore, only a single electron
can be bound in this confining potential since the energy of a second
electron will be in the continuum due to Coulomb repulsion. We have
shown that the spin-dependent interaction between a single propagating
electron and the weakly bound electron electron can induce entanglement
between them, giving rise to a two-electron quantum gate. In this
scenario, the flying qubit is realised as the spin of the propagating
electron and the static qubit is the spin of the bound electron. A
possible realisation of such a system is a clean semiconductor quantum
wire in which the propagating electron is injected through a single-electron
turnstile \cite{hu04} and the bound electron is trapped in a
shallow potential well along the wire, controlled by a gate electrode. 

The two-electron system may be modelled by the effective Hamiltonian\begin{equation}
H=-\sum_{i=1}^{2}\left[\frac{\hbar^{2}}{2m}\frac{\partial^{2}}{\partial x_{i}^{2}}+
v(x_{i})\right]+V(x_{1},x_{2}),\label{hamilt2}\end{equation}
where $m$ is the effective mass of an electron in the lowest conduction
miniband, $v(x)$ is an effective one-electron potential and $V(x_{1},x_{2})$
is an effective two-electron potential. This effective hamiltonian
accurately describes the system provided: (i) confinement in the transverse
dimensions is sufficintly large and the kinetic energy sufficiently
low that only the lowest miniband is occupied, (ii) the lowest transverse
mode is non-degenerate, (iii) the energy scale is sufficiently low
that non-parabolicity is negligible and (iv) the change in effective
potential $v(x)$ is sufficiently slow that coupling to higher minibands
is negligible. 
The effective potential in Eq.~(\ref{hamilt2})
is generic in that it may be explicitly induced, using surface gates,
or implicitly by an expansion in the transverse dimesions of the quantum
wire, or a combination of both. It may also have contributions from
remote charge centres or defects or other remote gates. The sources
of this confining potential are unimportant, and even the condition
that potential is slowly varying may be relaxed, e.g., by use of a
very narrow nanoscale gate. However, the effective potential well
must be sufficiently weak that to bind only a single electron, though
may have more single-electron bound states. We may regard this system
as an open quantum dot with Coulomb blockade precluding further electrons
from being bound. 
Such a system can show
exotic behaviour similar to the Kondo effect observed in more conventional
quantum-dot systems with high confining barriers \cite{goldhaber98}
and this behaviour has also been related to the conductance anomalies
referred to earlier and considered previously by the present authors
\cite{rrj00,rrj03} and others \cite{others}.

For the two-electron case we will show that the scattering of the
flying qubit from the static qubit can induce entanglement in a controlled
fashion and may thus be regarded as a candidate for realising a general
two-qubit gate and explicitly demonstrating exchange of quantum information
between a static qubit and a flying qubit. Consider an unentangled
state in which the quantisation axis is chosen to be in the direction
of the propagating electron spin and the bound-electron spin is in
some general state on the Bloch sphere, i.e. $\cos(\vartheta/2)|\downarrow>+e^{i\phi}\sin(\vartheta/2)|\uparrow>$$.$
We may write the antisymmetrised incoming scattering states for the
two electrons as\begin{eqnarray}
\Psi_{in} & =c\Psi_{\uparrow\downarrow}^{+}+e^{i\phi}s\Psi_{\uparrow\uparrow}^{+}.\label{inpsi}\end{eqnarray}
where $c=\cos(\vartheta/2)$, $s=\sin(\vartheta/2)$ and\[
\Psi_{\sigma\sigma'}^{\pm}=\left|\begin{array}{cc}
e^{\pm ikx_{1}}\chi_{1\sigma} & \: e^{\pm ikx_{2}}\chi_{2\sigma}\\
\psi_{b}(x_{1})\chi_{1\sigma'} & \:\psi_{b}(x_{2})\chi_{2\sigma'}\end{array}\right|.\]
Here $\psi_{b}(x)$ is the ground-state wavefunction of the bound
electron and the injected electron has quasimomentum $k$ and spinor
$\chi_{\uparrow}$. After scattering, the propagating electron will
be reflected or transmitted and, asymptotically, will have the same
magnitude of momentum, $k$, leaving the bound electron again in its
ground state, $\psi_{b}$, provided that the initial energy of the
incoming electron is smaller than the energy separation, $\Delta E$,
from the the bound-state to the next allowable state. In cases where
there is only one bound state, $\Delta E$ is the ionisation energy,
otherwise it is the threshold energy for inelastic scattering via
intra-dot transitions. For elastic scattering, the reflected and transited
part of the asymptotic states are\begin{eqnarray}
 & \Psi_{out}^{-} & =c(r_{nsf}\Psi_{\uparrow\downarrow}^{-}+r_{sf}\Psi_{\downarrow\uparrow}^{-})+e^{i\phi}sr_{\uparrow\uparrow}\Psi_{\uparrow\uparrow}^{-}\nonumber \\
 & \Psi_{out}^{+} & =c(t_{nsf}\Psi_{\uparrow\downarrow}^{+}+t_{sf}\Psi_{\downarrow\uparrow}^{+})+e^{i\phi}st_{\uparrow\uparrow}\Psi_{\uparrow\uparrow}^{+}.\label{outpsi}\end{eqnarray}

We see that both the reflected and transmitted waves show spin entanglement
after scattering provided $\cos(\vartheta/2)$ and the amplitudes
for spin-flip and non-spin-flip scattering, $r_{sf},\, r_{nsf},\, t_{sf}$
and $t_{nsf}$ are non-zero. Furthermore, fully entangled states occur
when $\vartheta=0$ and $|r_{sf}|=|r_{nsf}|$ or $|t_{sf}|=|t_{nsf}|$.
Although it is clear that the interaction between electrons will induce
entanglement, it is not obvious that this can be controlled or indeed
that maximum entanglement can be achieved. Full entanglement seems
plausible for the following reasons. Writing the asymptotic states
in the basis of spin eigenstates and comparing with Eqs.~(\ref{outpsi})
we see directly that\begin{equation}
t_{nsf}=\frac{t_{T}+t_{S}}{2},\: t_{sf}=\frac{t_{T}-t_{S}}{2}\:\textrm{and}\: t_{\uparrow\uparrow}=t_{T}.\label{eq:tnsf}\end{equation}
 Now we know that this two electron system has at least a singlet
resonance ($|t_{S}|=1)$ at some energy for which the triplet state
is off resonance $(|t_{T}|\ll1)$. It follows that at this resonance,
$|t_{sf}|\approx|t_{nsf}|\approx\frac{1}{2}$ and the state is close
to being fully entangled when spins are initially antiparallel ($\cos(\vartheta/2)=1)$.
Similarly, in reflection, $|r_{S}|\approx0$, $|r_{T}|\approx1$ and
$|r_{sf}|\approx|r_{nsf}|\approx\frac{1}{2}$, which is also close
to being fully entangled. Note that at such a resonance, the propagating
electron has approximately equal probability $(\frac{1}{2})$ of being
either transmitted or reflected. The precise condition for full entanglement
in transmission is $\vartheta=0,$ $|t_{T}+t_{S}|=|t_{T}-t_{S}|$
and this is satisfied when the complex numbers $t_{S}$ and $t_{T}$
are at right angles in the Argand diagram, i.e. $\delta_{S}-\delta_{T}=(2n+1)\frac{\pi}{2}$,
where $\delta$ is the phase shift due to scattering and $n$ is an
integer. 

\begin{figure}
\includegraphics[%
  width=55mm]{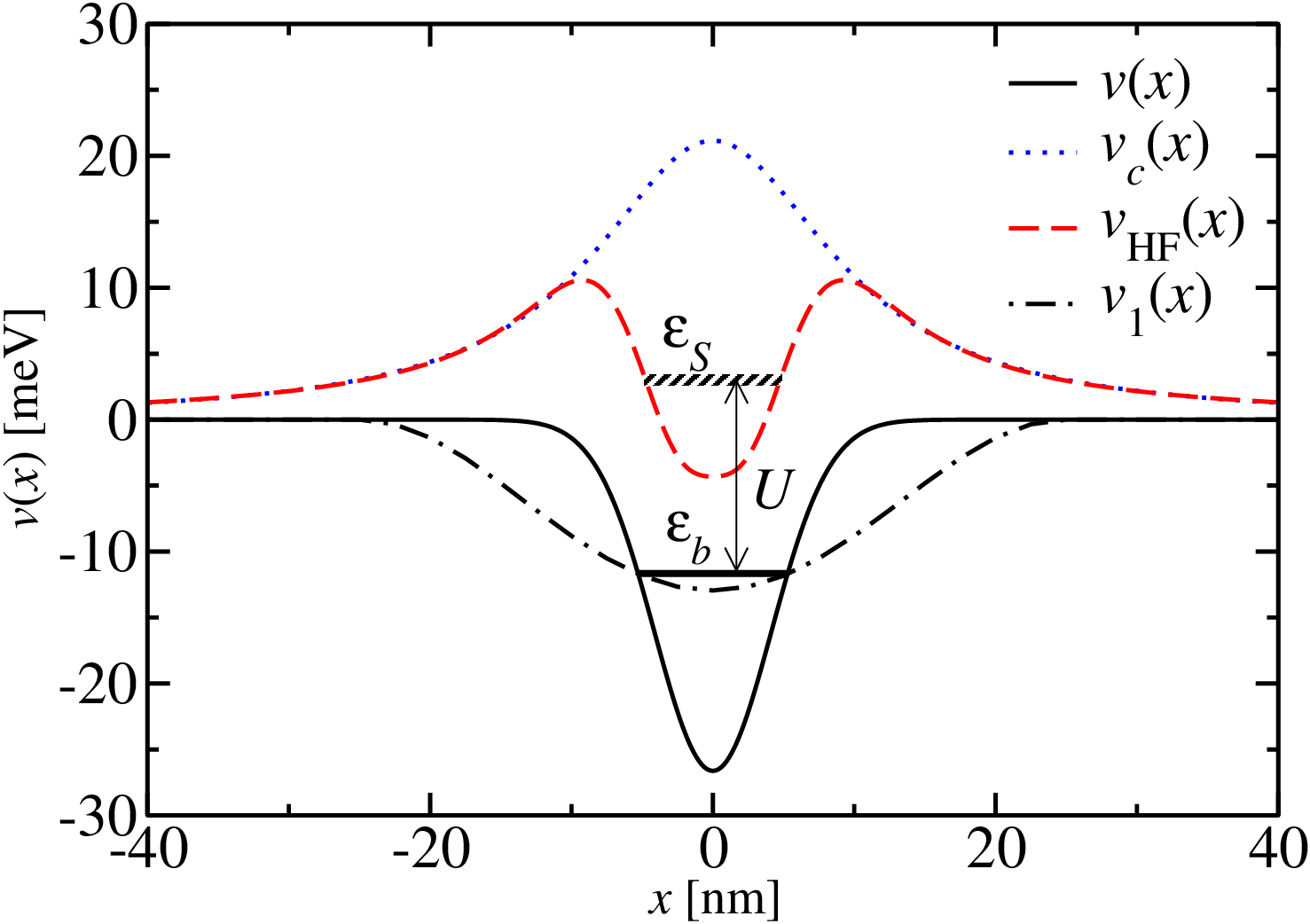}

\caption{\label{cap:Fig1}(color online) Shallow potential well $v(x)$ (full
line) and corresponding Hartree-Fock potential seen by a second electron
of opposite spin(dashed line). The single resonant bound state, $\epsilon_{b}$,
and quasi-bound singet state, $\epsilon_{S}$, within the double barrier
structure are also indicated. The dotted line is the Coulomb repuslion
energy due to bound electron and the dashed-dotted line represents
shallow potential well $v_{1}(x)$ corresponding for the results from
Fig.~\ref{cap:Fig3}.}
\end{figure}

When the states are not fully entangled, a measure of their degree
of entanglement is given by the concurrence \cite{wootters98} which,
for the pure states considered here, is defined as\begin{equation}
C=\frac{2}{\langle\Psi|\Psi\rangle}\left|\begin{array}{cc}
\langle\alpha\alpha|\Psi\rangle & \langle\alpha\beta|\Psi\rangle\\
\langle\beta\alpha|\Psi\rangle & \langle\beta\beta|\Psi\rangle\end{array}\right|,\label{eq:concurrence}\end{equation}
where $|\alpha\rangle$and $|\beta\rangle$ are orthogonal qubit base
states in any representation. For transmitted electrons the asymptotic
spin-concurrence after scattering is thus, from Eqs.~(\ref{outpsi}),
(\ref{eq:tnsf}), and (\ref{eq:concurrence})\begin{eqnarray}
 & C_{t}=\frac{2c(T\, T_{nsf})^{1/2}}{c^{2}\left[T_{sf}+T_{nsf}\right]+s^{2}
T_{\uparrow\uparrow}}\,\,\,\,\,\,\,\,\,\,\,\,\,\,\,\,\,\,\,\,\,\,\,\,\,\,\,\,\,\,\,\,\,\,\,\,\,\,\,\,\nonumber \\
 & \,\,\,\,\,=\frac{2c\large[(T_{T}+T_{S})^{2}-4T_{T}T_{S}\cos^{2}
(\delta_{T}-\delta_{S})\large]^{1/2}}{T_{T}+T_{S}+s^{2}(T_{T}-T_{S})} & ,\label{eq:ct}\end{eqnarray}
where $T_{\lambda}=|t_{\lambda}|^{2}$ for corresponding labels $\lambda$.
Similarly, $|r_{T}+r_{S}|=|r_{T}-r_{S}|$ for full entanglement in
reflection with antiparallel spins initially. We again see that full
entanglement is plausible at $\vartheta=0$ for energies for which
either pure singlet or pure triplet states are near a resonance since
we know, for example, that the phase shift for resonant singlet scattering
changes rapidly as we sweep through the resonance energy, whereas
the phase shift for the triplet varies only slowly provided the overlap
of the resonance widths is small. Thus, provided that singlet and
triplet energies are not too close in energy, there will be some energy
for which $\cos(\delta_{T}-\delta_{S})=0$ in Eq.~(\ref{eq:ct}).
We also note from Eq.~(\ref{eq:ct}) that the concurrence approaches
unity when either $T_{S}\gg$$T_{T}$ or $T_{T}\gg$$T_{S}$. This
is simply due to the fact that one of the spin channels (singlet or
triplet) is 'filtered out' leaving the other channel which is fully
entangled.

\begin{figure}
\includegraphics[%
  width=55mm]{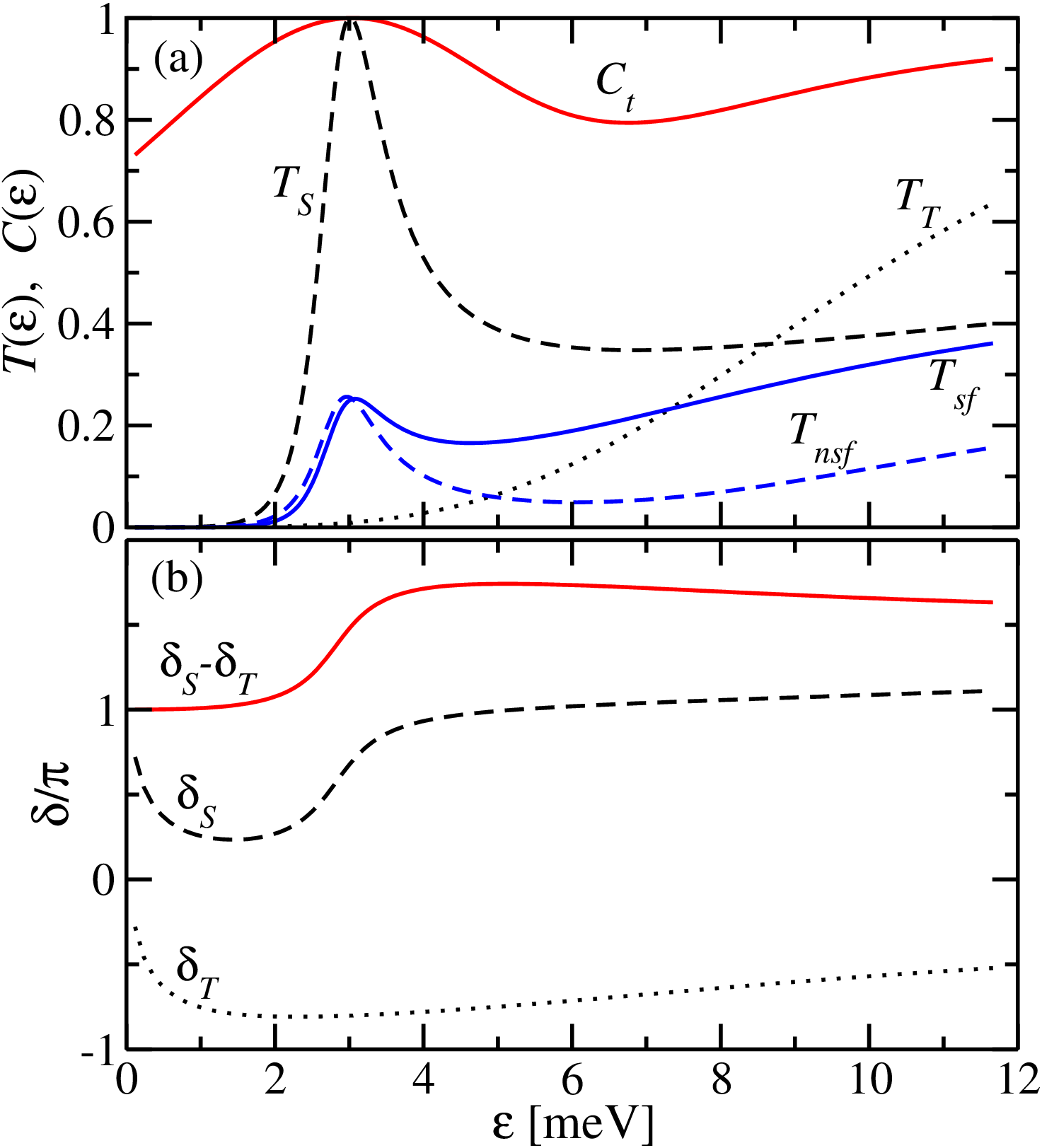}

\caption{\label{cap:Fig2}(color online) (a) Singlet and triplet transmission
probalility ($T_{S},$~$T_{T}$), spin-flip and non-spin-flip transmission
probability ($T_{sf}$, $T_{nsf}$) and corresponding concurrence
$C_{t}$ for confining potential $v(x)$ from Fig.~\ref{cap:Fig1}.
(b) Phase shifts corresponding to transmission probabilities in (a).}
\end{figure}

Further illustration of this behaviour is seen by solving the scattering
problem explicitly for specific cases. 
Numerical solutions for symmetrised (singlet) or antisymmetrised
(triplet) orbital states  
yield directly the complex amplitudes $t_{S},\, r_{S},\, t_{T},\, r_{T}$
from which the amplitudes for spin-flip and non-spin-flip scattering
may be calculated using Eq.~(\ref{eq:tnsf}). We have obtained results
for GaAs quantum wires with an effective mass $m=0.067m_{0}$, a wire
width of $10$~nm giving an energy separation of $125$~meV between
the lowest and first excited transverse modes, and an effective Coulomb
interaction $V(x_{1},x_{2})$ given by integrating the bare 3D Coulomb
interaction over the lowest transverse mode. The shallow effective
potential, $v(x)$, is first chosen such that there is only a single
one-electron bound state at energy $\epsilon_{b}=-12$~meV with a
well depth of $26$~meV and a width of $\sim20$~nm (Fig.~\ref{cap:Fig1}).
As described in previous work \cite{rrj00,rrj03}, the bound electron
has a long-range Coulomb interaction with the propagating electron
and, when combined with the well potential, gives rise to a double
barrier structure which has a singlet resonance energy at approximately
$\epsilon_{s}\sim\epsilon_{b}+U$, where $\epsilon_{b}$ is the energy
of the lowest bound state and $U=15$~meV is the Coulomb matrix element
for two electrons of opposite spin occupying this state. This is also
shown in Fig.~\ref{cap:Fig1} where we have plotted the Hartree-Fock
potential due to the bound electron, $v_{HF}(x)$, i.e. the self-consistent
potential seen by the propagating electron in the 'frozen' potential
due to the bound electron of opposite spin. 

In Fig.~\ref{cap:Fig2} we plot the singlet and triplet transmission
and reflection probabilities, showing a single maximum of unity for
the transmission. We also plot concurrence, which is very close to
the corresponding singlet resonance, occuring when the spin-flip probabilities
are equal and approximately $\frac{1}{4}$ in both transmission and
reflection. The phase angle of the singlet changes rapidly with energy
through the resonance, whereas the triplet resonance is fairly flat.
Since the total change in singlet phase angle is somewhat in excess
of $\pi/2$, there is point where the singlet-triplet phase difference
is precisely $\pi/2$ and the transmitted state is fully entangled.
The behviour is similar in reflection. Note, however, that at low
energy, the difference in phase angle for singlet and triplet tends
to $\pi$ in transmission and the limiting concurrence is non-zero,
whereas in reflection the limiting behaviour is zero phase shift and
concurrence. This can be understood when we consider that at low-energy
the total transmission probability is very small and hence this somewhat
unexpected behaviour results from a very improbable transmission event.
When this does occur, the spin-flip process dominates since the incoming
up-spin electron simply displaces the down-spin electron due to Coulomb
repulsion. Neglecting the non-spin-flip process we see, from Eq.~{[}\ref{eq:ct}{]}
that $t_{S}\approx-t_{T}$ , i.e. a phase difference of $\pi$. Actually,
the limiting non-spin-flip process, though small, is not negligible,
as can be seen from the finite concurrence. This shows a limiting
value of around $0.7$ giving $T_{nsf}/T_{sf}\sim0.1$. In reflection,
the non-spin-flip process is dominant at low energy by the same argument
(as seen explicitly in the plot) and hence the phase difference tends
to zero as does the concurrence.

\begin{figure}
\includegraphics[%
  clip,
  width=55mm]{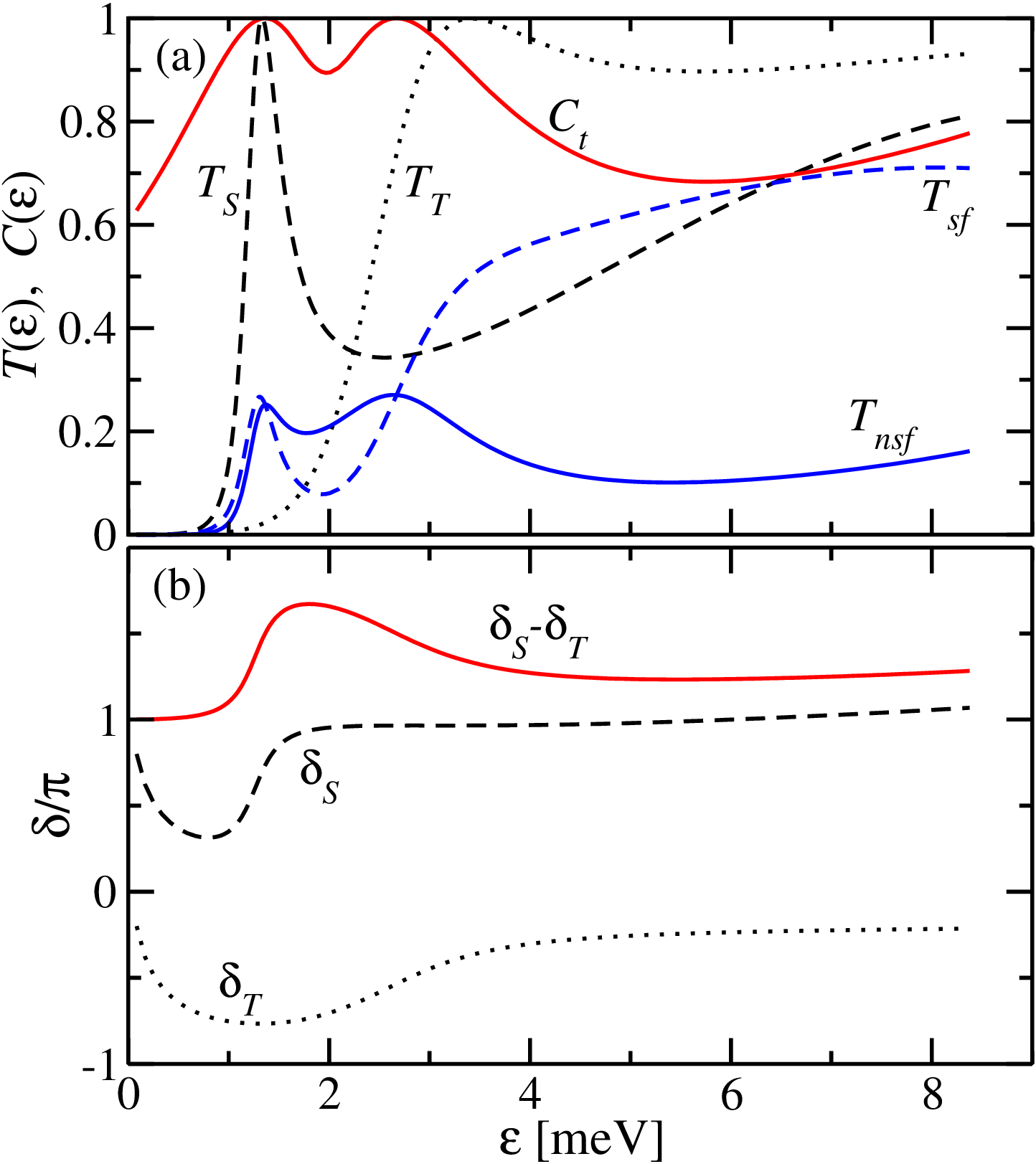}

\caption{\label{cap:Fig3}(color online) (a) Singlet and triplet transmission
probalility ($T_{S},$~$T_{T}$), spin-flip and non-spin-flip transmission
probability ($T_{sf}$, $T_{nsf}$) and corresponding concurrence
$C_{t}$ for confining potential $v_{1}(x)$ from Fig.~\ref{cap:Fig1}.
(b) Phase shifts corresponding to transmission probabilities in (a).}
\end{figure}

In Fig.~\ref{cap:Fig3} we show results for a shallow potential well
of depth $12$~meV and width $\sim40$~nm. With these parameters
there are two single-electron bound states at energies $-8$~meV
and $-10$~meV. This gives both a singlet and a triplet resonance,
the latter corresponding to one electron in the lowest bound state
and the other in the higher bound state which becomes a resonance
obeying Hund's rule under Coulomb repulsion, with a further singlet
resonance outside the energy window for elastic scattering. We see
that there are two unitary peaks of concurrence in transmission with
the second close to, but clearly discernable from, the peak of the
rather broad triplet resonance. We have shown in other examples, where
singlet and triplet resonances are very close, that the concurrence
does not always reach the unitary limit, since both singlet and triplet
phase shifts vary rapidly with energy with their difference not reaching
$\pi/2.$

In conclusion, we have shown that spin-entanglement and exchange of
quantum information occurs via the Coulomb interaction when a propagating
electron interacts Coulombically with a single bound electron in a
shallow potential well in a one-dimensional semiconducting quantum
wire. The degree of entanglement may be controlled by kinetic energy
of the incoming electron and the shape of the effective potential
well and unitary concurrence occurs near a singlet or triplet resonance.
Potential realisations of such a system are semiconductor quantum
wires and carbon nanotubes. 

A possible sequence of operations to demonstrate that entanglement
has been achieved would be as follows. Initialisation would consist
of first loading the open quantum dot with a single electron using
a turnstile injector with surface and back gates to control the shape
and depth of the potential well. A second electron is then injected
through the turnstile, incorporating a Zeeman quantum dot spin-filter
\cite{spinfilter} in a global magnetic field. The spin filter is
tuned such that only minority spins are resonant, resulting in a propagating
electron with opposite spin to the bound electron. Alternatively,
both static and propagating spins may have the same polarisation with
the spin of the bound electron flipped by a microwave $\pi$-pulse
prior to interaction. Any further spin rotation due to the global
magnetic field may then be accounted for explicitly. This would, of
course, depend on the group velocity of the injected electron which
may be controlled by the source drain bias, enablling the kinetic
energy of the incident electron to sweep the resonances. Measurement
of spin for the propagating electron after interaction could also
be done using a quantum-dot spin-filter in which a transmitted electron
would be detected by a single-electron transistor. The static spin
would then be inferred indirectly by injecting a second propagating
spin of known polarisation and correlating its measured spin after
interaction with that of the first propagating electron.

We thank Andrew Briggs, Crispin Barnes, Simon Benjamin, Mike Fearn,
George Giavaras, Daniel Gunlycke, David Pettifor and Sandu Popescu
for helpful discussions. Authors acknowledge support from the EU,
the UK MoD, the UK IRC on QIP (GR/S82176/01), and the Ministry of
Higher education, Science and Technology of Slovenia under grant Pl-0044.


\begin{thebibliography}{10}
\bibitem{Loss98}D. Loss and D. P. DiVincenzo, Phys. Rev. A \textbf{57}(1), 120 (1998).
\bibitem{Elzerman04}J.M. Elzerman, R. Hanson, L.H. Willems van Beveren, B. Witkamp,
L.M.K. Vandersypen, and L.P. Kouwenhoven, Nature \textbf{430} 431
(2004). 
\bibitem{Hanson03}R. Hanson, B. Witkamp, L. M. K. Vandersypen, L. H. Willems van Beveren, 
J. M. Elzerman, and L. P. Kouwenhoven, Phys. Rev. Lett. \textbf{91}
196802 (2003).
\bibitem{Hanson05} R. Hanson, L. H. Willems van Beveren, I. T. Vink, J. M. Elzerman, W. J. M. Naber, 
F. H. L. Koppens, L. P. Kouwenhoven, and L. M. K. Vandersypen, Phys.
Rev. Lett. \textbf{94}, 196802 (2005).
\bibitem{Petta04}J.R. Petta, A.C. Johnson, A. Yacoby, C.M. Marcus, M.P. Hanson, and
A.C. Gossard A C, cond-mat/0412048.
\bibitem{burkard04}G. Burkard and D. Loss, Phys. Rev. Lett. \textbf{91}, 087903 (2004).
\bibitem{hu04} Xuedong Hu, and S. Das Sarma, Phys. Rev. B \textbf{69}, 115312 (2004).
\bibitem{costa01}A.T. Costa, Jr. and S. Bose, Phys. Rev. Lett. \textbf{87}, 277901
(2001).
\bibitem{oliver02}W.~D.~Oliver, F.~Yamaguchi, and Y.~Yamamoto Phys. Rev. Lett. \textbf{88},
037901 (2002).
\bibitem{Barnes00} C. H. W. Barnes, J. M. Shilton, and A. M. Robinson, Phys. Rev. B
\textbf{62}, 8410 (2000).
\bibitem{1bound}L D Landau and E M Lifshitz, Quantum Mechanics: Non-Relativistic Theory,
Volume 3, Pergamon (1977).

\bibitem{goldhaber98}D.~Goldhaber-Gordon, H. Shtrikman, D. Mahalu, D. Abusch-Magder, U.
Meirav, and M.A. Kastner, Nature \textbf{391}, 156 (1998).
\bibitem{rrj00}T. Rejec, A. Ram{\v s}ak, and J. H. Jefferson, Phys. Rev. B \textbf{62},
12985 (2000); J. Phys.: Condens. Matter \textbf{12}, L233 (2000).
\bibitem{rrj03}T. Rejec, A. Ram{\v s}ak, and J. H. Jefferson, Phys. Rev. B \textbf{67},
075311 (2003).
\bibitem{others}Y. Meir, K. Hirose, and N. S. Wingreen, Phys. Rev. Lett. \textbf{89},
196802 (2002); O.P. Sushkov, Phys. Rev. B \textbf{67}, 195318 (2003);
A.A. Starikov, I.I. Yakimenko, and K.-F. Berggren, i\emph{bid}. \textbf{67},
235319 (2003); P.S. Cornaglia, C.A. Balseiro, and M. Avignon, \emph{ibid.}
\textbf{71}, 024432 (2005).
\bibitem{wootters98}W. K. Wootters, Phys. Rev. Lett. \textbf{80}, 2245 (1998).
\bibitem{spinfilter}P. Recher, E.V. Sukhorukov, and D. Loss, Phys. Rev. Lett. 85, 1962
(2000).

\end{thebibliography}
\end{document}